\documentclass{appolb}
\usepackage{graphicx}
\usepackage[abs]{overpic}
\usepackage[dvipsnames]{xcolor}
\usepackage{tikz}

\begin{document}
\title{Pion and kaon pair production in double gap events in ALICE Run 3
\thanks{Presented at ``Diffraction and Low-$x$ 2024'', Trabia (Palermo, Italy), 
September 8-14, 2024.}%
}
\author{Rainer Schicker, for the ALICE Collaboration
  \address{Physikalisches Institut, Universität Heidelberg, Im Neuenheimer Feld 226, Heidelberg, Germany}
\\[3mm]
}
\maketitle
\begin{abstract}
  The ALICE detector at the LHC has undergone a major upgrade in the long
  shutdown 2019-2021 to be able to take data at much-increased rates in
  Runs 3 and 4. The upgrades of the detector systems used for analysing
  double gap events are described, and the improvement in data taking
  capability for such double gap events is presented.
  
\end{abstract}
  
\section{Introduction}
In the long shutdown LS2 of 2019-2021, the ALICE experiment at the LHC was
upgraded in order to take data at much-increased rates in Runs 3 (2022-2025)
and 4 (2029-2032) \cite{ALICEup}. The readout was changed to continuous mode,
with data collected in proton-proton collisions at rates up to \mbox{650 kHz,}
and subsequently being processed to identify the events of interest. The
selected events are permanently stored for future processing. All data taken
in heavy-ion collisions at up to 50 kHz are permanently stored, and available
for later analysis.

\section{The detector upgrades}
The Time-Projection-Chamber (TPC) is not only the main tracking detector of
the ALICE experiment but also provides information on the specific ionisation
energy loss dE/dx for particle identification. The ionisation electrons drift in
the electric field of 400 V/cm to the endplates of the cylindrical field cage.
The ionisation charges were amplified in Runs 1 \mbox{(2009-2013)} and
2 (2015-2018)
by multiwire proportional \mbox{chambers (MWPCs),} with the backflow of positive
ions controlled by gating grids. The effective rate was limited in
Runs 1 and 2 to about 1 kHz for Pb-Pb and about \mbox{3 kHz} for pp collisions.
In order to increase the rate capability to about \mbox{50 kHz} in Pb-Pb,
the MWPCs were replaced by chambers with a stack of 4 Gas
Electron Multipliers (GEMs) with continuous readout and synchronous
\mbox{data processing \cite{TPCup}.}

In Runs 1 and 2, the readout rate of the Inner Tracking System (ITS) was
limited to 1 kHz. The new ITS system implemented for Run 3 features improved
resolution, less material and faster readout. The new system consists
of 7 layers of Monolithic Active Pixel Sensors, with a  spatial resolution
\mbox{(r$\phi$ x z)} improved from (11 x 100) $\mu$m$^{2}$ of the old ITS
to (5 x 5) $\mu$m$^{2}$. The readout rate of this new ITS system is 100 kHz
in Pb-Pb \cite{ITSup}.

With these TPC and ITS  upgrades, data have been taken so far in pp
collisions at rates up to 650 kHz, and tests for rates up to 1 MHz are planned.

Outside of the central barrel pseudorapidity range $|\eta| < $ 0.9, a Fast
Interaction Trigger (FIT) system was built for Run 3 to provide precise
collision time information for time-of-flight-based particle information, for
online luminosity monitoring and measuring forward multiplicity \cite{FITup}.
This FIT system consists of the following detectors:

\begin{itemize}
\item {\bf FT0:} Two Cherenkov arrays for minimum bias triggering,
  for determining the collision time and for vertex position calculation.
\item {\bf FV0:} A large scintillator array consisting of five rings and eight
  sectors is positioned on the opposite side of the ALICE muon spectrometer.
  In conjunction with FT0, it is used for centrality and event plane
  determination in heavy-ion collisions.
\item {\bf FDD} (Forward Diffractive Detector): This double-sided scintillator
  array is essential for tagging diffractive events by establishing
  rapidity gaps in the event.
\end{itemize}
\begin{table}[h]
\begin{center}
  \begin{tabular}{p{0.18\textwidth}|p{0.12\textwidth}|p{0.12\textwidth}|p{0.10\textwidth}}
  Detector & $\eta_{min}$ & $\eta_{max}$ & z [cm]\\
  \hline
  FDD-A & 4.8 & 6.3 & 1696.0\\
  FT0-A & 3.5 & 4.9 & 334.6\\
  FV0   & 2.2 & 5.1 & 320.8 \\
  FT0-C & -3.3 & -2.1 & -84.3 \\
  FDD-C & -7.0 & -4.9 & -1956.6\\
\end{tabular}  
\caption{The pseudorapidity range and the z-position of the FIT subsystems.}
\label{table1}
\end{center}
\end{table}

The parameters of the different FIT subsystems are shown in Table \ref{table1}. 


Events with particle production measured in the ALICE central barrel
and absence of signals in the FIT systems are tagged as double gap events.

\section{The computing system upgrade}
After the upgrade of the ALICE detector systems in LS2,
the raw data flow to the data acquisition system  increased a
hundredfold, up to 3.5 TB/s. A new Online/Offline Computing system, O$^{2}$,
was developed to cope with this challenge \cite{O2up}. The O$^{2}$ system
incorporates continuous readout of most subdetectors, data compression using
partial synchronous  reconstruction and calibration, and the sharing of common
computing resources during and for asynchronous reconstruction after data
taking. The reconstructed data are written to disk while the raw data are
discarded. The O$^{2}$ architecture consists
of two major computing layers, the First Level Processors (FLPs) and
the Event Processing Nodes (EPNs). Both layers are highly heterogeneous,
with specialized acquisition cards embedding FPGAs on the FLPs, and
GPUs on the EPNs. The raw data rate of 3.5 TB/s is reduced by the
FLP layer to $\sim$ 635 GB/s  and by the EPN layer to \mbox{100 GB/s.}
This compressed data stream is stored and, after later asynchronous
processing, distributed to the Tier 0 and Tier 1 analysis facilities.

  \begin{figure}[h]
\begin{center}    
\begin{overpic}[width=.68\textwidth]{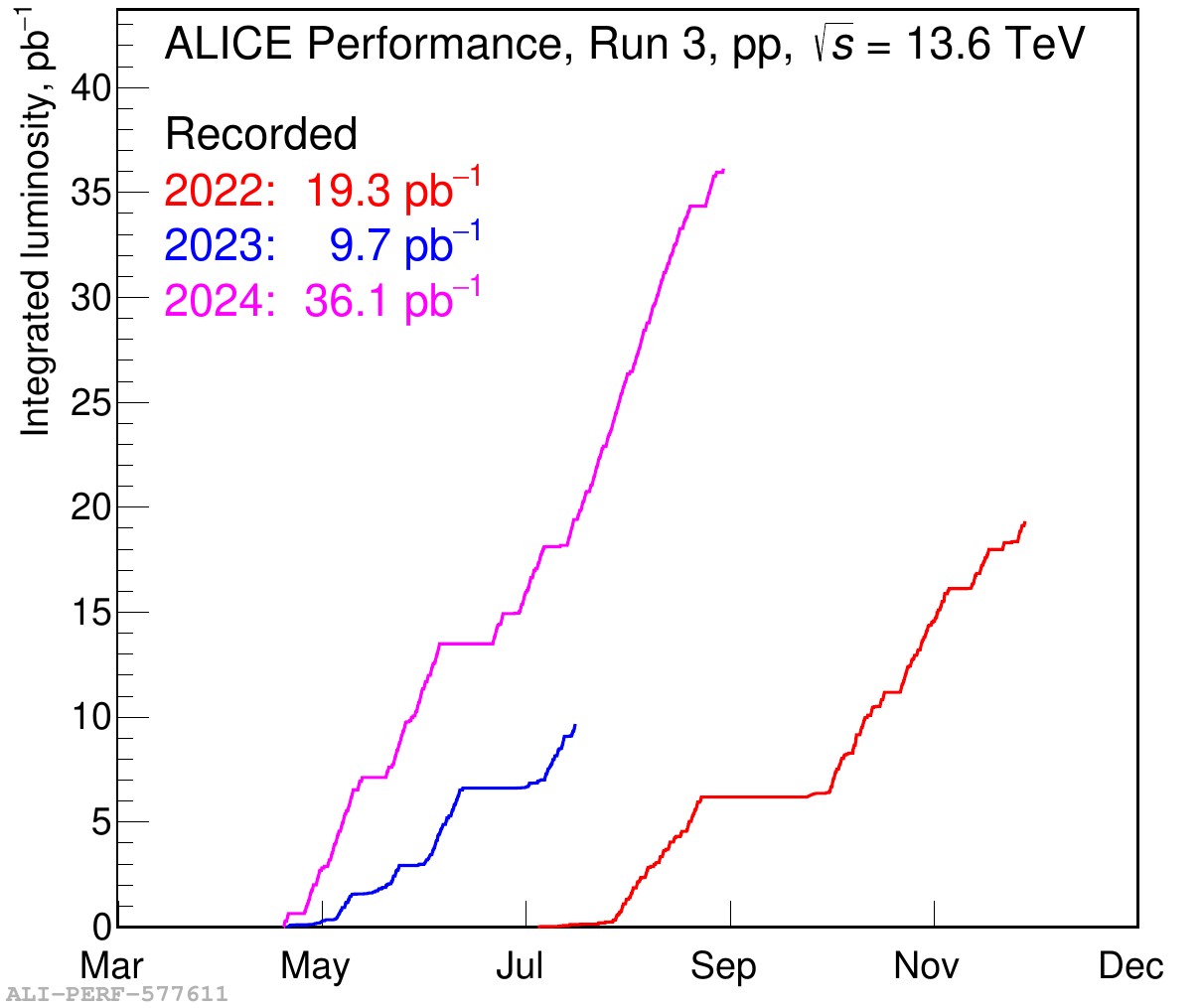}
    \put(127.8,94.){\color{CarnationPink}\linethickness{0.2mm}\vector(-1,0){12.0}}
    \put(127.8,94.){\color{CarnationPink}\linethickness{0.2mm}\vector(1,0){12.0}}

    \put(181.,43.0){\color{red}\linethickness{0.2mm}\vector(-1,0){12.}}
    \put(181.,43.0){\color{red}\linethickness{0.2mm}\vector(1,0){12.}}


    \put(140.4,128.){\color{CarnationPink}\linethickness{0.2mm}\vector(0,1){33.4}}
    \put(140.4,128.){\color{CarnationPink}\linethickness{0.2mm}\vector(0,-1){33.4}}

    \put(192.8,64.){\color{red}\linethickness{0.2mm}\vector(0,1){20.}}
    \put(192.8,64.){\color{red}\linethickness{0.2mm}\vector(0,-1){20.}}

    \put(114.0,85.0){\scriptsize\textcolor{CarnationPink}{$\!\sim\!$ 6 $\!$wks}}
    
    \put(168.,33.){\scriptsize \textcolor{red}{$\!\sim\!$ 6 $\!$wks}}

    \put(144.0,112.0){\rotatebox{90}{\scriptsize\textcolor{CarnationPink}{$\!\sim\!$ 16 $\!$pb$^{\!-\!1}$}}}
    
    \put(196.0,46.0){\rotatebox{90}{\scriptsize\textcolor{red}{ $\!\sim\!$ 10 $\!$pb$^{\!-\!1}$}}}

    \put(167.,61.0){\rotatebox{64}{\scriptsize\textcolor{red}{2022}}}
    \put(114.,120.0){\rotatebox{70}{\scriptsize\textcolor{CarnationPink}{2024}}}

\end{overpic}
\end{center}
\caption{The Run 3 ALICE data sample amounts to about 10 pb$^{-1}$ per 6 weeks
  of data taking in 2022, and to about 16 pb$^{-1}$ per 6 weeks in 2024.}
\label{fig:lumi}
  \end{figure}
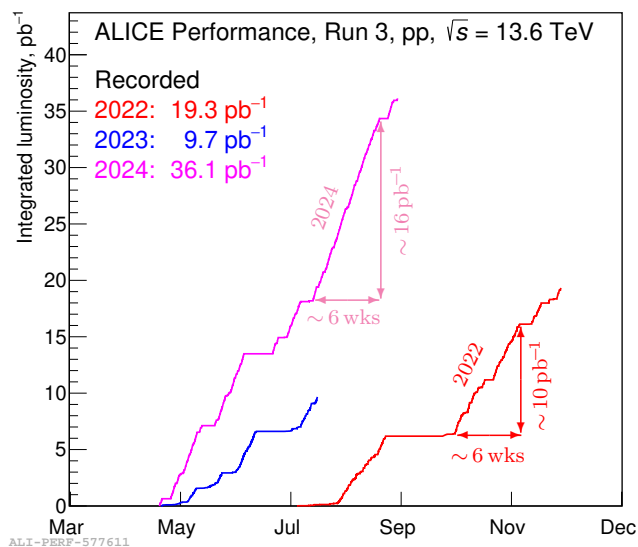

This upgrade of the ALICE experiment resulted in a considerably increased
data taking capability. The double gap data sample collected in Run 2 amounts
to about 10 pb$^{-1}$. As shown in Fig. \ref{fig:lumi}, the ALICE data sample
was about 10 pb$^{-1}$ per 6 weeks of data taking in 2022, and about 16 pb$^{-1}$
per 6 weeks of data taking in 2024.

\section{Central diffractive production at the LHC}

Central diffractive events are characterized by particle production at
midrapidity, and by rapidity gaps up to the rapidity of the beam particle
or its remnants. The ALICE experiment is ideally suited for studying such
events due to the excellent global tracking in the central barrel by the
combined information of TPC and ITS and the superb particle identification
of the TPC. The absence of particles at forward/backward
rapidities can be established by requiring no signals in the FIT detector
systems. In the following, only events with two particles at
midrapidity are shown. 

       \begin{figure}[h]
    \begin{overpic}[width=.52\textwidth]{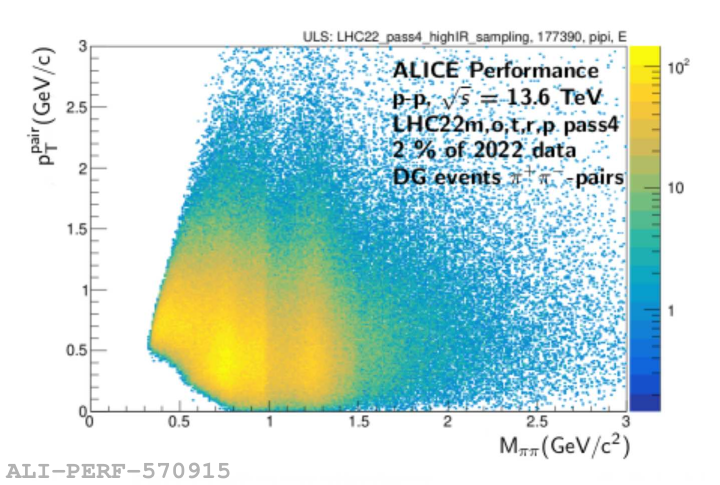}
    \end{overpic}
    \hspace{-.1cm}
    \begin{overpic}[width=.52\textwidth]{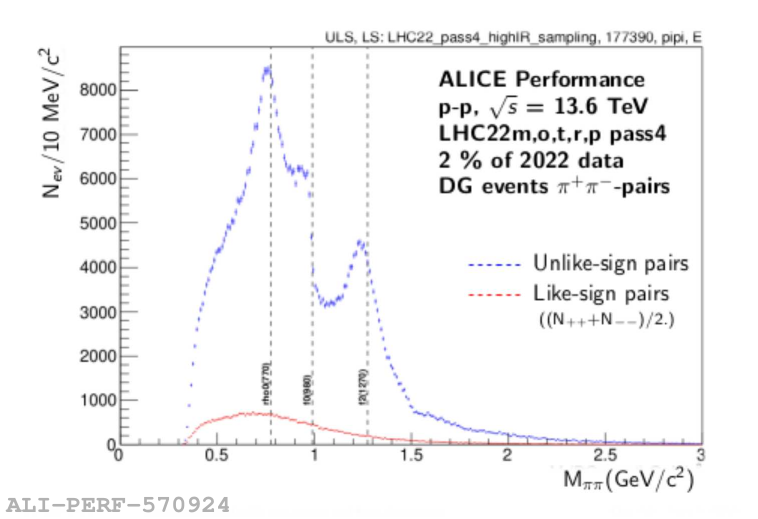}
\put(62.4,28.){\rotatebox{90}{\colorbox{White}{\makebox(16.,1.0){\tiny{$\rho$(770)}}}}}
\put(73.2,26.){\rotatebox{90}{\colorbox{White}{\makebox(18.,1.0){\tiny{$f_{0}$(980)}}}}}
\put(86.6,22.){\rotatebox{90}{\colorbox{White}{\makebox(21.,1.0){\tiny{$f_{2}$(1270)}}}}}
    \end{overpic}
\caption{Transverse momentum vs. invariant mass of unlike-sign pion pairs on
  the left, invariant mass distribution for like and unlike-sign pion pairs
  on the right.}
    \label{fig:dg_pion}
  \end{figure}

\begin{figure}[h]
\includegraphics[width=.53\textwidth]{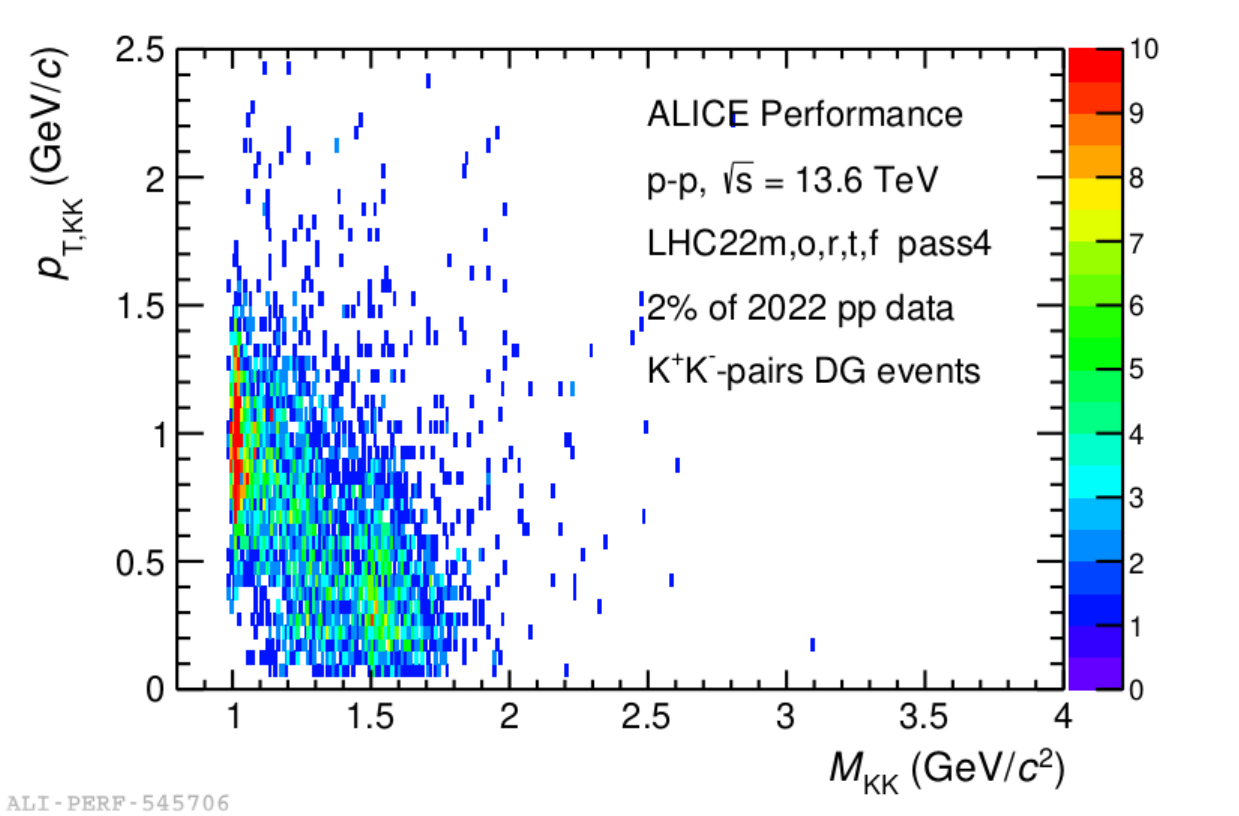}
\includegraphics[width=.54\textwidth]{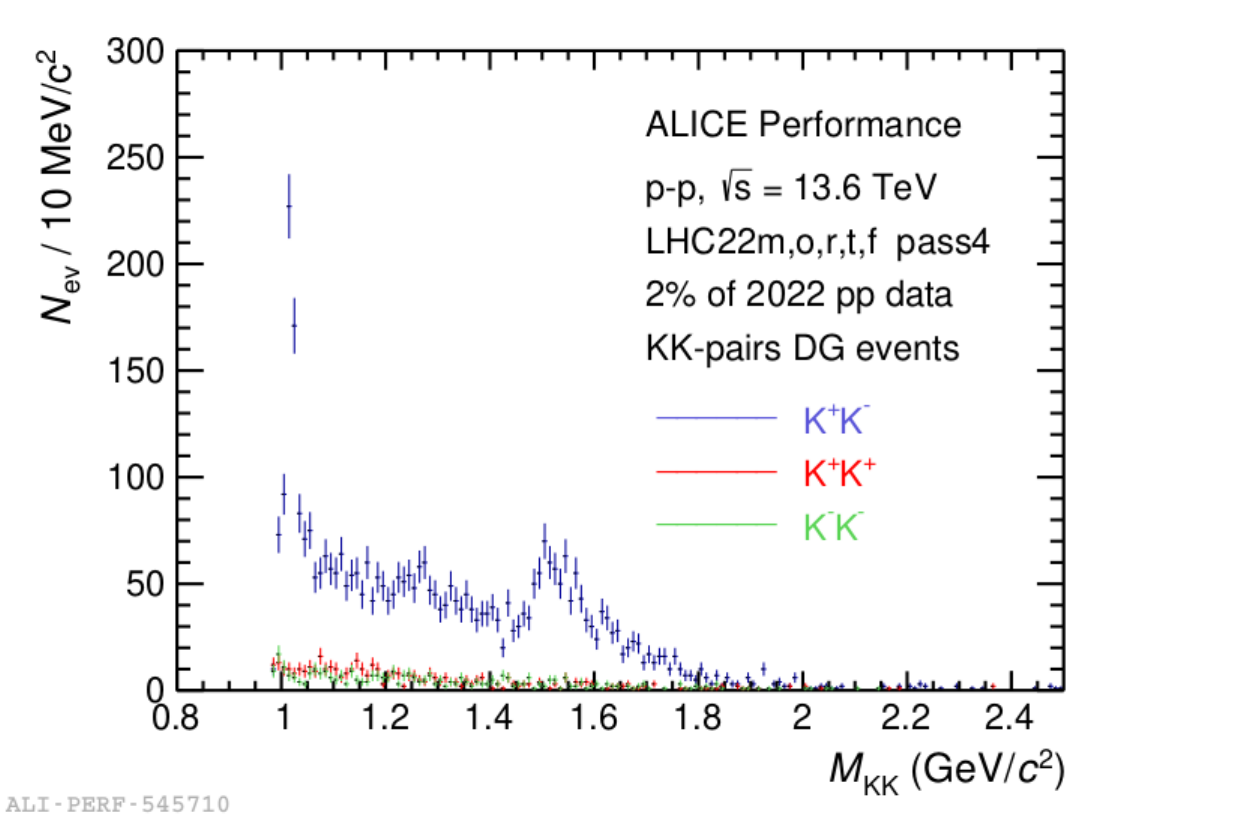}
\caption{Transverse momentum vs. invariant mass of unlike-sign kaon pairs on
  the left, invariant mass distribution for like and unlike-sign kaon pairs
  on the right.}
\label{fig:dg_kaon}
  \end{figure}

    In Fig. \ref{fig:dg_pion} on the left, the transverse momentum $p_{\rm T}$ of
    unlike-sign pion pairs is shown vs. the pair mass $M_{\pi\pi}$, reflecting
    the ALICE pair acceptance.
    Integrating this distribution over the pair $p_{\rm T}$ yields the pion pair
    invariant mass distribution shown on the right of Fig. \ref{fig:dg_pion}.
    The corresponding distributions of kaon pairs are shown
    in Fig. \ref{fig:dg_kaon}.

    The invariant pair mass spectra in both the pion and kaon sectors
    shown in Figs. \ref{fig:dg_pion} and \ref{fig:dg_kaon} exhibit
    clear resonance structures. The study of resonance production
    in double gap events is of high interest since such events
    are dominated at LHC energies by Pomeron-Pomeron fusion.
    In Quantumchromodynamics (QCD), the Pomeron is understood to be
    a colour-neutral exchange of an even number of gluons.
    Pomeron-Pomeron fusion events are hence thought to produce a gluon
    rich environment with quark degrees of freedom largely missing
    in the initial state. Such a medium is the ideal place
    to hunt for exotic hadronic states such as hybrids and glueballs.
    A first step in such a hunt is, however, the analysis of the data
    for known $q\bar{q}$ states.

    \section{A model of $q\bar{q}$ bound states}

    A unified framework formulated by Godfrey  and Isgur for calculating
    $q\bar{q}$ bound states of light and heavy mesons relies on a
    relativistic potential \cite{Isgur}

    \begin{equation}
      {\small V(\mathbf p,\mathbf r) = H^{conf} + H^{so} + H^{hyp} + H_{A}}
\label{eq:Isgur}
    \end{equation}

    with $H^{conf}$ the confining potential,  $H^{so}$ the spin-orbit interaction,
    $H^{hyp}$ the hyperfine interaction and $H_{A}$ the annihilation interaction.
    
    The solutions of Eq. \ref{eq:Isgur} are grouped according to their quantum
    numbers $J^{PC}$, and associated to ground and excited states in
    spectroscopic notation $n\:^{2S+1}L_{J}$ in the different flavour sectors.

    \begin{table}[h]
      \caption{Isoscalar mesons with hidden strangeness.}
      \label{table2}
\begin{center}
      \begin{tabular}{|p{0.10\textwidth}|p{0.15\textwidth}|p{0.20\textwidth}|p{0.20\textwidth}|}
\hline
\small{PDG} & \small{$J^{PC}$ (G-I)} & \small{$n\:^{2S+1}L_{J}$ (G-I)}  & \small{mass (G-I)}   \\
\hline
\small{$\phi$} & 1$^{--}$ & \small{$1^{3}S_{1}$} &\small{1020 MeV/c$^{2}$}    \\
\small{$f_{2}^{'}$} & 2$^{++}$ & \small{$1^{3}P_{2}$} &\small{1530 MeV/c$^{2}$} \\
\small{$\phi_{3}$} & 3$^{--}$ &  \small{$1^{3}D_{3}$} &\small{1900 MeV/c$^{2}$} \\
\small{$f_{4}^{'}$} & 4$^{++}$ & \small{$1^{3}F_{4}$} &\small{2200 MeV/c$^{2}$} \\
\small{$\phi_{5}$} & 5$^{--}$ & \small{$1^{3}G_{5}$} &\small{2470 MeV/c$^{2}$} \\
\hline
      \end{tabular}
      \end{center}
\end{table}

    In Table \ref{table2}, the S,P,D,F and G-wave solutions of the
    Godfrey-Isgur model in the isoscalar sector with hidden strangeness are
    listed. The S-wave solution with a calculated mass of 1020 MeV/c$^{2}$ is
    identified with the well known $\phi$(1019). The P-wave solution with a
    calculated mass of 1530 MeV/c$^{2}$ is identified with the well known
    $f_{2}^{'}$  of mass 1525 MeV/c$^{2}$. Both the $\phi$(1019) and the
    $f_{2}^{'}$(1525) are clearly visible in the kaon pair mass spectrum shown
    in Fig. \ref{fig:dg_kaon}. The D-wave solution of the Godfrey-Isgur model
    with  a calculated mass of 1900 MeV/c$^{2}$ is identified with
    the $\phi_{3}$(1850) listed by the Particle Data \mbox{Group \cite{pdg}.}
    The kaon pair invariant mass spectrum shown in Fig. \ref{fig:dg_kaon}
    presently shows very little statistics in the mass range above
    \mbox{1700 MeV/c$^{2}$.} The data sample shown in Fig. \ref{fig:dg_kaon}
    represents, however, only about 2\% of the data sample recorded in 2022.
    There are no states listed by the Particle Data Group corresponding to
    the F-wave ($J^{PC} = 4^{++}$) and G-wave ($J^{PC} = 5^{--}$) solutions
    of the Godfrey-Isgur model, hence the last two entries of Table \ref{table2}
    labeled as $f_{4}^{'}$ and $\phi_{5}$ represent terra incognita.

\section{Summary and outlook}

The ALICE experiment is taking data at unprecedented rates in \mbox{Run 3}
after a major upgrade in LS2. First analyses of kaon pair production
in double gap events show clear evidence for strangeonia states
$\phi$(1019) and $f_{2}^{'}$(1525). There is a 50 times larger  data
sample for analysis from data taking 2022/2023, and an even
larger sample available from data taking 2024.  The
analysis presented here can be extended to $(u,d)\bar{s}$ kaonia  and
    $(\bar{u},\bar{d})s$ antikaonia states  by studying  $\pi$K pairs.

\section{Acknowledgements}

This work is supported by the German Federal Ministry of Education and
Research under reference 05P24VHA.

\bibliographystyle{unsrt}
\bibliography{schicker}

\end{document}